# The Interpretation of Magnetisation and Entropy Jumps in the Flux-line Lattice

A.I.M. Rae*, E.M. Forgan* & R.A. Doyle[†]

*School of Physics and Astronomy, University of Birmingham,
Edgbaston, Birmingham B15 2TT.
[†]IRC in Superconductivity, University of Cambridge, West Cambridge Site,
Madingley Road, CB3 0HE

**In the HTC superconductor BSCCO, local measurements of magnetic field in the mixed state show sharp changes as a function of applied field or temperature, which have been interpreted as evidence for a first-order flux-lattice melting (or sublimation) transition [1]. The entropy associated with such a transition was calculated from the magnetisation jumps using a Clausius-Clapeyron-type relationship and was found to increase with temperature up to a value of nearly $1.5k_B$ per flux-line per CuO layer near $T_c$. This is considerably larger than would be expected, even if the fluid phase consisted of completely dissociated 'pancake' vortices. We re-examine this calculation and show that the total change in entropy associated with the transition is actually over $4k_B$ per flux-line per CuO layer. However, we also show that the major portion of this entropy can be attributed to the cores of the extra flux lines formed in the transition and that the residue is consistent with that expected for the disorder associated with melting or sublimation, thereby resolving an important controversy.**

Experimental measurements [1] of the local $B$ field at the surface of a HTC plate-like crystal, subject to an applied magnetic field perpendicular to the plane of the plate have been performed using microscopic Hall sensors, based on a two-dimensional electron gas, which have a spatial resolution of about 3μm. Sharp jumps ($\Delta B$) in this field were observed at particular values of the temperature and magnetic flux density. These have been interpreted as evidence for a first-order phase transition, probably associated with the loss of crystalline order in the flux-line lattice as is indicated by neutron diffraction measurements[2] and numerical simulations[3]. The entropy change, $\Delta S$, associated with this transition has been calculated[1] from the following expression

$$\Delta S = -\frac{d\phi_0}{B_m} \frac{dB_m}{dT_m} \Delta B \text{ per flux-line per CuO layer} \tag{1}$$

where $d$ is the separation between CuO planes in the crystal, $B_m$ and $T_m$ are the values of the magnetic flux density at the transition and $\phi_0$ is the flux quantum ($h/2e$). The resulting values of $\Delta S(T)$ are shown in Fig. 1 where they are shown to rise to a value of about $1.5k_B$ close to $T_c$. This is considerably larger than would be expected, even from a complete loss of three-dimensional order at the transition, as has been confirmed by Monte-Carlo simulations[4] which predict $\Delta S \sim 0.5 k_B$. The purpose of this letter is to suggest a re-interpretation of the experimental data that provides a physical reason for the large entropy change.

The standard expression for the entropy based on the Clausius-Clapeyron relation at a first-order magnetic transition[5] is

$$\Delta S = -\frac{d\phi_0}{B_m} \mu_0 \frac{dH_m}{dT_m} \Delta M \tag{2}$$

which differs from (1) in two respects. First, $\Delta M$ replaces $\Delta B$. Clearly if two adjacent sections of the crystal are in different phases, they are in equilibrium at the same value of $H$ and are separated by a current sheet carrying a current of magnitude $\Delta M$ per unit length. Far from the top and bottom edges of such a sheet, all fields are parallel to it and clearly $\Delta M = \Delta B$. However at the surface of the crystal where the measurements are made, $\Delta M = 2\Delta B$. This is most easily seen by re-considering the field at the mid line between the upper and lower surfaces, perpendicular to the current sheet. The component of $\Delta B$ ($=\Delta M$) parallel to the sheet has identical contributions from the two halves of the current sheet which lie above and below this mid line. If one half of the specimen, including the current sheet, is removed, the mid-line becomes a surface and $\Delta B$ is halved so that it now equals $\Delta M/2$. The effect of this correction is to double $\Delta S$, making its maximum value about $3.0 k_B$, which is even more different from that predicted from a simple melting model.

The second difference between (1) and (2) is that $dB_m/dT_m$ is replaced by $\mu_0 dH_m/dT_m$. From elementary electromagnetism, the difference these expressions is $\mu_0 dM_m/dT_m$. There are a number of expressions for the magnetisation of a type II superconductor depending on the value of the applied field relative to $H_{c1}$ and $H_{c2}$[6], but in all cases the temperature dependence is proportional to that of $H_{c1}$ apart from possible logarithmic terms which we ignore. We therefore put

$$\frac{\partial M}{\partial T} = -\alpha \frac{\partial H_{c1}}{\partial T} \tag{3}$$

where $\alpha$ is a numerical constant whose value can be deduced from the experimental magnetisation curves[1] as 0.4  Assuming that the temperature dependence of $H_{c1}$ is given by

$$H_{c1}(T) = H_{c10}\left(1 - \frac{T^2}{T_c^2}\right) \tag{4}$$

and using the standard relations for a type II superconductor, we get

$$\frac{\partial H_m}{\partial T_m} = \frac{1}{\mu_0}\frac{\partial B_m}{\partial T_m} - \alpha \frac{2T_m}{T_c^2} H_{c10} = \frac{1}{\mu_0}\frac{\partial B_m}{\partial T_m} - \alpha \frac{\phi_0 T_m}{2\pi\lambda_0^2 T_c^2}\ln\kappa$$

(5)

where $\lambda_0$ is the London penetration depth at $T = 0$ and $\kappa$ is the Ginsburg-Landau parameter. Fig. 1 includes the entropy change calculated, including both the corrections discussed so far. We see that it now reaches a value of about $4.2 k_B$ near $T_c$, an order of magnitude greater than that expected on the basis of loss of positional order.

We now proceed to discuss an alternative source of entropy that we believe is implicit in this situation. We assume that the transition consists of the disordering of a set of vortices and that these objects still exist in the fluid phase. (The argument that follows applies equally to line and pancake vortices) In particular, we assume that the flux associated with each vortex is a flux quantum and that the cores of the vortices are unaffected by the transition; this is consistent with the fact that the transition takes place well below $H_{c2}$ so that there is no significant overlap of the cores. It follows directly that a positive change in magnetisation in a finite-sized sample inevitably accompanied by a corresponding increase in the number of vortices per unit area. However the core of each vortex has an entropy associated with it, so the total entropy per unit volume must increase quite apart from any entropy associated with the disorder. Moreover, this entropy change *per unit volume* will be proportional to $\Delta M$ so that, when expressed *per flux line* the entropy change $\Delta S$ will diverge at low fields (assuming $\Delta M$ constant) and hence near $T_c$.

We can make a quantitative estimate of this entropy change from the standard expression for the free energy of an isolated flux line in a high-$\kappa$ superconductor [6]

$$F_l = \phi_0 H_{c1} = \frac{\phi_0^2 \ln\kappa}{4\pi\mu_0 \lambda(T)^2}$$

(6)

So that the corresponding entropy, $S_l$, is given by

$$S_l = -\frac{\partial F_l}{\partial T} = \frac{\phi_0^2 T \ln\kappa}{2\pi\mu_0 \lambda_0^2 T_c^2} \quad \text{again assuming that} \quad \lambda(T) = \lambda_0\left[1 - \left(\frac{T}{T_c}\right)^2\right]^{-\frac{1}{2}}$$

(7)

Hence, given that the contribution to the magnetisation from an isolated flux line is $M_l = \phi_0/\mu_0$,

$$\frac{S_l}{M_l} = \frac{\mu_0 S_l}{\phi_0} = \frac{\phi_0 T \ln\kappa}{\pi\lambda_0^2 T_c^2}$$

(8)

If this were the only entropy change associated with the transition, then this quantity would be equal to $-\mu_0 \frac{\partial H_m}{\partial T_m}$. Moreover, we see that the form of this expression is identical to the contribution to the entropy from the magnetisation change considered earlier - apart from the factor of $-\alpha$. This means that, since $\alpha > -1$, this new correction will more than cancel out the previous one and lead to an overall reduction in the nett entropy change associated with the melting disorder *per se*. We have evaluated (8) with the same parameters as used earlier and subtracted this contribution from the total entropy change; this is also included in Fig. 1. We see that it is nearly independent of temperature over most of the temperature range and varies from $0.1 k_B$ to about $0.6 k_B$ per flux line per layer which is consistent with the values expected for a melting transition and obtained from Monte-Carlo calculations[4]. The precise value of the residual $\Delta S$ is sensitive to the value of the penetration depth and its temperature dependence, but, although the value of $\lambda_0$ used above (140nm) was deliberately chosen to make this quantity small near $T_c$, it is well within the accepted range of experimental values[7].

We conclude that the experimental data on the phase transition in the flux line lattice is consistent with the loss of three-dimensional translational order. The simplest model of this is that the fluid phase consists of a set of mobile disordered pancake vortices. Any alternative model of this phase would have to have the same ratio of superconducting to normal state volumes if it is to reproduce the observed total entropy change.

We are grateful to Eli Zeldov and Archie Campbell for useful discussions
and to Bill Powell who many years ago taught one of us (AIMR) about the importance of understanding **B** and **H**.

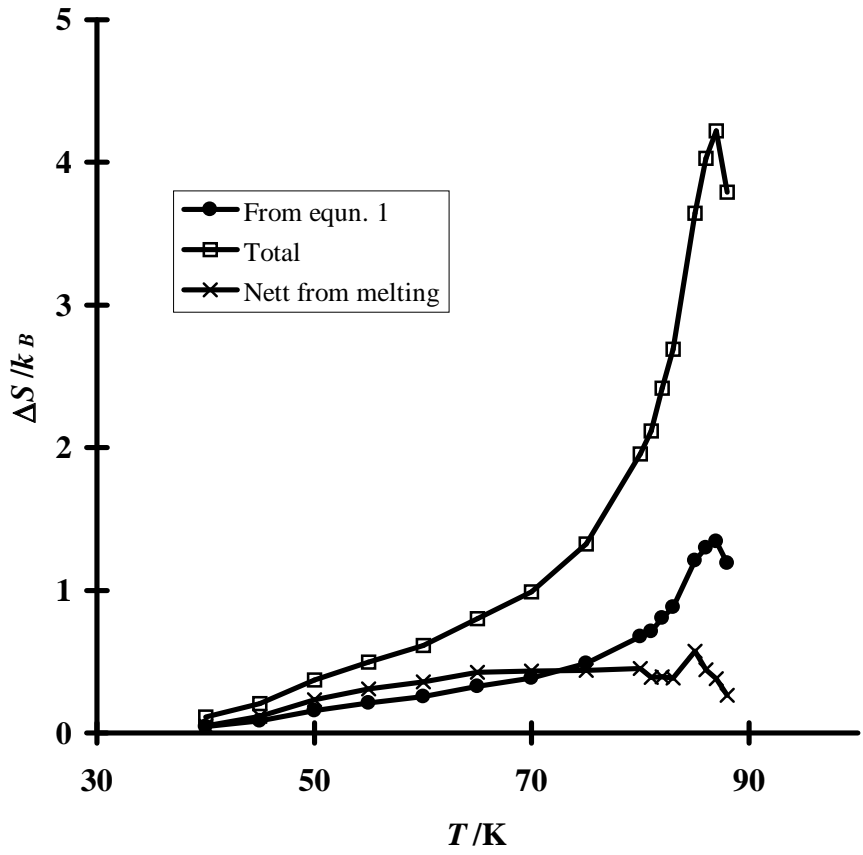

Fig. 1

Figure Legend

The entropy change per flux line per CuO layer as a function of temperature calculated from the form of the Clausius-Clapeyron relation in ref2, the total entropy change calculated by the method set out in this paper and the nett entropy change associated with the melting as such.